\documentclass[aps,prb,twocolumn,showpacs,superscriptaddress]{revtex4}

\usepackage{graphicx}

\begin{document}

\title{Stacking-order dependent transport properties of trilayer graphene}

\author{S.~H.~Jhang} \altaffiliation{e-mail:sungho.jhang@physik.uni-regensburg.de}
\affiliation{Institute of Experimental and Applied Physics, University of Regensburg, 93040
Regensburg, Germany}

\author{M.~F.~Craciun}
\affiliation{Centre for Graphene Science, CEMPS University of Exeter, EX4 4QF Exeter, United Kingdom}

\author{S.~Schmidmeier}
\affiliation{Institute of Experimental and Applied Physics, University of Regensburg, 93040
Regensburg, Germany}

\author{S.~Tokumitsu}
\affiliation{Department of Applied Physics, University of Tokyo, Tokyo 113-8656, Japan}

\author{S.~Russo}
\affiliation{Centre for Graphene Science, CEMPS University of Exeter, EX4 4QF Exeter, United Kingdom}

\author{M.~Yamamoto}
\affiliation{Department of Applied Physics, University of Tokyo, Tokyo 113-8656, Japan}

\author{Y.~Skourski}
\affiliation{Dresden High Magnetic Field Laboratory,
Helmholtz-Zentrum Dresden-Rossendorf, 01314 Dresden, Germany}

\author{J.~Wosnitza}
\affiliation{Dresden High Magnetic Field Laboratory,
Helmholtz-Zentrum Dresden-Rossendorf, 01314 Dresden, Germany}

\author{S.~Tarucha}
\affiliation{Department of Applied Physics, University of Tokyo, Tokyo 113-8656, Japan}

\author{J.~Eroms}
\affiliation{Institute of Experimental and Applied Physics, University of Regensburg, 93040
Regensburg, Germany}

\author{C.~Strunk}
\affiliation{Institute of Experimental and Applied Physics, University of Regensburg, 93040
Regensburg, Germany}

\begin{abstract}
We report markedly different transport properties of ABA- and ABC-stacked trilayer graphenes.
Our experiments in double-gated trilayer devices provide evidence that a perpendicular electric field opens an energy gap in the ABC trilayer,
while it causes the increase of a band overlap in the ABA trilayer.
In a perpendicular magnetic field, the ABA trilayer develops quantum Hall plateaus at filling factors of $\nu =$~2, 4, 6... with a step of $\Delta\nu =$~2,
whereas the inversion symmetric ABC trilayer exhibits plateaus at $\nu =$~6 and 10 with 4-fold spin and valley degeneracy.
\end{abstract}


\maketitle

The unique chiral nature of low-energy quasiparticles in graphene, characterized by a Berry phase $J\pi$ with linear and parabolic dispersion for monolayer ($J=1$) and bilayer ($J=2$) graphene respectively, results in unusual quantum Hall effects (QHE). \cite{Novoselov,Zhang,Novoselov1}
The Landau-level (LL) energy in a perpendicular magnetic field $B$, given by $E_n \propto \sqrt{Bn}$ for monolayer
and $E_n \propto B \sqrt{n(n-1)}$ for bilayer graphene, shows $J$-fold degenerate LLs at zero energy, with integer $n$ being the LL index.
The existence of $J$-fold degenerate zero-energy LLs, combined with 4-fold spin and valley degeneracy in each LL,
explains the unusual sequence of quantum Hall states observed at filling factor sequences $\nu=\pm2, \pm6, \pm10...$ for monolayer,\cite{Novoselov,Zhang}
and $\nu=\pm4, \pm8, \pm12...$ for bilayer graphene.\cite{Novoselov1}
The bilayer graphene is further distinguished from the gapless monolayer by a tunable energy gap, induced by breaking
the inversion symmetry of the two layers in a perpendicular electric field.\cite{Castro,Oostinga,Wang}

In few-layer graphene (FLG), the stacking order offers an extra degree of freedom.
Indeed, the electronic structure and the Landau level spectrum differ significantly depending on the stacking order in FLG.\cite{Guinea,Min,Koshino,Koshino1,Avetisyan}
For instance, Bernal (ABA)-stacked trilayer exhibits an electric-field tunable band overlap,\cite{Craciun,Koshino3} while rhombohedral (ABC)-stacked trilayer is predicted to present a tunable band gap.\cite{Guinea,Koshino1,Avetisyan} To date, no direct evidence of the electric-field and stacking-order dependent transport properties has yet been reported in double-gated devices.
In the simplest tight-binding model that includes only the nearest intra- and inter-layer hopping parameters $\gamma_0$ and $\gamma_1$ (Fig.~1(b)),
the Landau level spectrum of the ABA trilayer can be viewed as a superposition of $\sqrt{B}$-dependent monolayer-like LLs and $B$-dependent bilayer-like LLs (Fig.~1(c)).
On the other hand, LLs of the ABC trilayer (Fig.~1(d)) are given by $E_n \propto B^{3/2} \sqrt{n(n-1)(n-2)}$ with Berry's phase $3\pi$.\cite{Guinea,Min}
Despite the substantial difference in the LL spectrum, 3-fold degenerate zero-energy LLs with 4-fold spin and valley degeneracy
are expected to result in QHE plateaus at filling factor sequences $\nu = \pm6, \pm10, \pm14...$ for the trilayer graphene independently of the stacking order.\cite{Guinea,Min,Ezawa,Koshino2}
However, the lack of inversion symmetry in ABA trilayer may lead to a broken valley degeneracy,
while the valley degeneracy of LLs is guaranteed in the inversion-symmetric ABC trilayer.\cite{Koshino2}
Here, we report stacking-dependent transport properties of double-gated trilayer graphene, combined with Raman spectroscopy.
We show that the effects of applied electric and magnetic fields on the ABC-stacked trilayers are strikingly different from those on the ABA-stacked trilayers.

\begin{figure}[b]
\includegraphics[width=8.6cm]{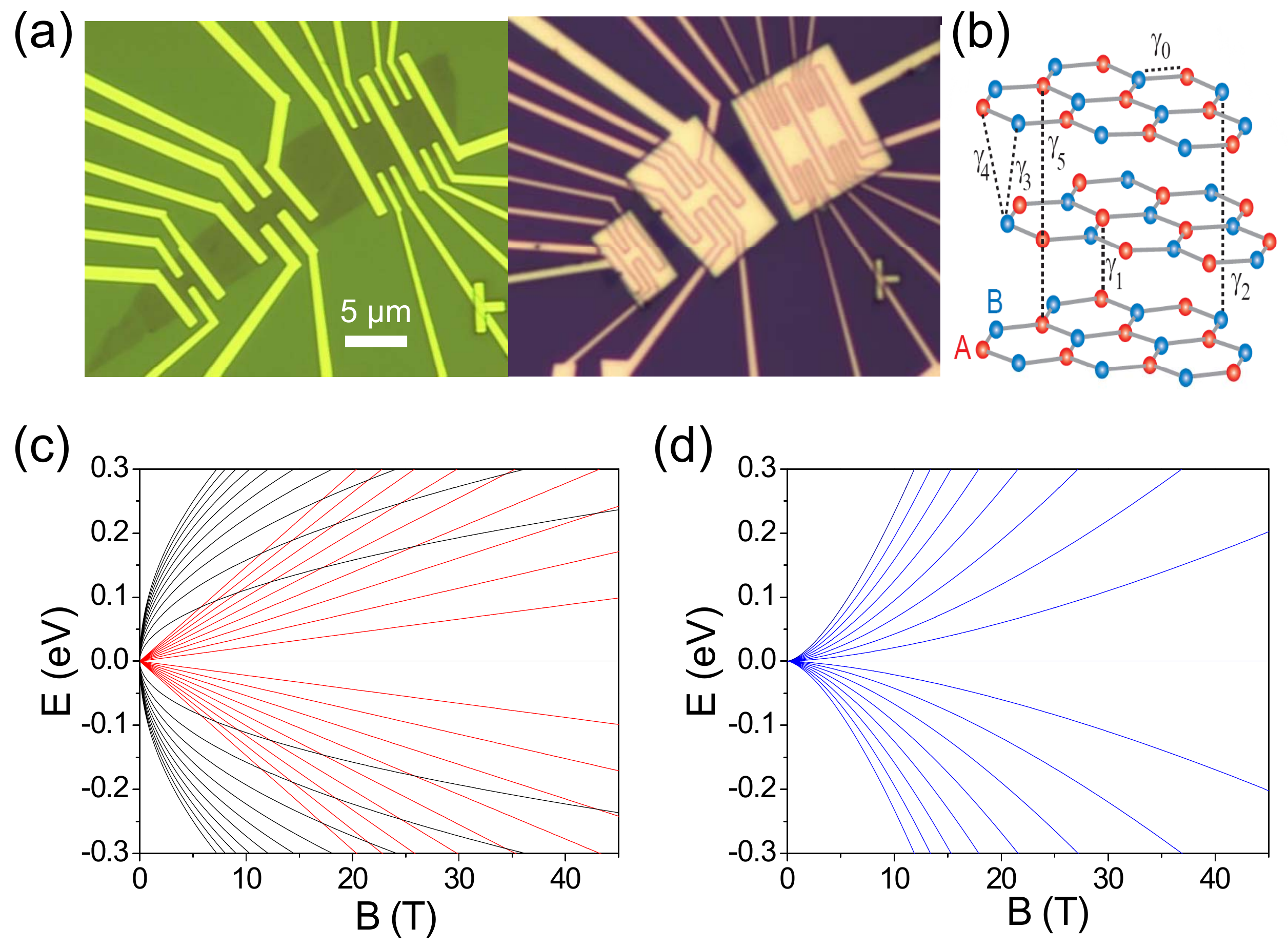}
\caption{(Color online) (a) Optical images of a typical device before and after top-gate deposition. (b) The structure of ABA-stacked trilayer graphene. (c) Landau levels of ABA-stacked trilayer graphene shown up to $n$ = 10, calculated with $\gamma_0 = 3$~eV and $\gamma_1 =$ 400~meV.
(d) Landau levels of ABC-stacked trilayer graphene shown up to $n$ = 10.
}\label{fig:LLs}
\end{figure}

Our experiments have been performed on single- and double-gated trilayer graphene devices\cite{Russo,Craciun1} prepared by exfoliating graphite on Si/SiO$_{2}$ substrates.
The heavily conductive Si was used as a back gate and the thickness of the SiO$_{2}$ layer was 285~nm.
Ti/Au electrodes and top gates (SiO$_{2}$/Ti/Au) were defined on top of the graphene flakes by electron-beam lithography (Fig.~1(a)).
The dc magneto-transport properties were studied at liquid-Helium temperatures in pulsed perpendicular magnetic fields of up to 50~T.
The magnetic-field pulse resulted from the discharge of a large capacitor bank with a capacitance of 30~mF and a voltage up to 20~kV
and lasted typically $\simeq500$~ms.

\begin{figure}[t]
\centering
\includegraphics[width=8.6cm]{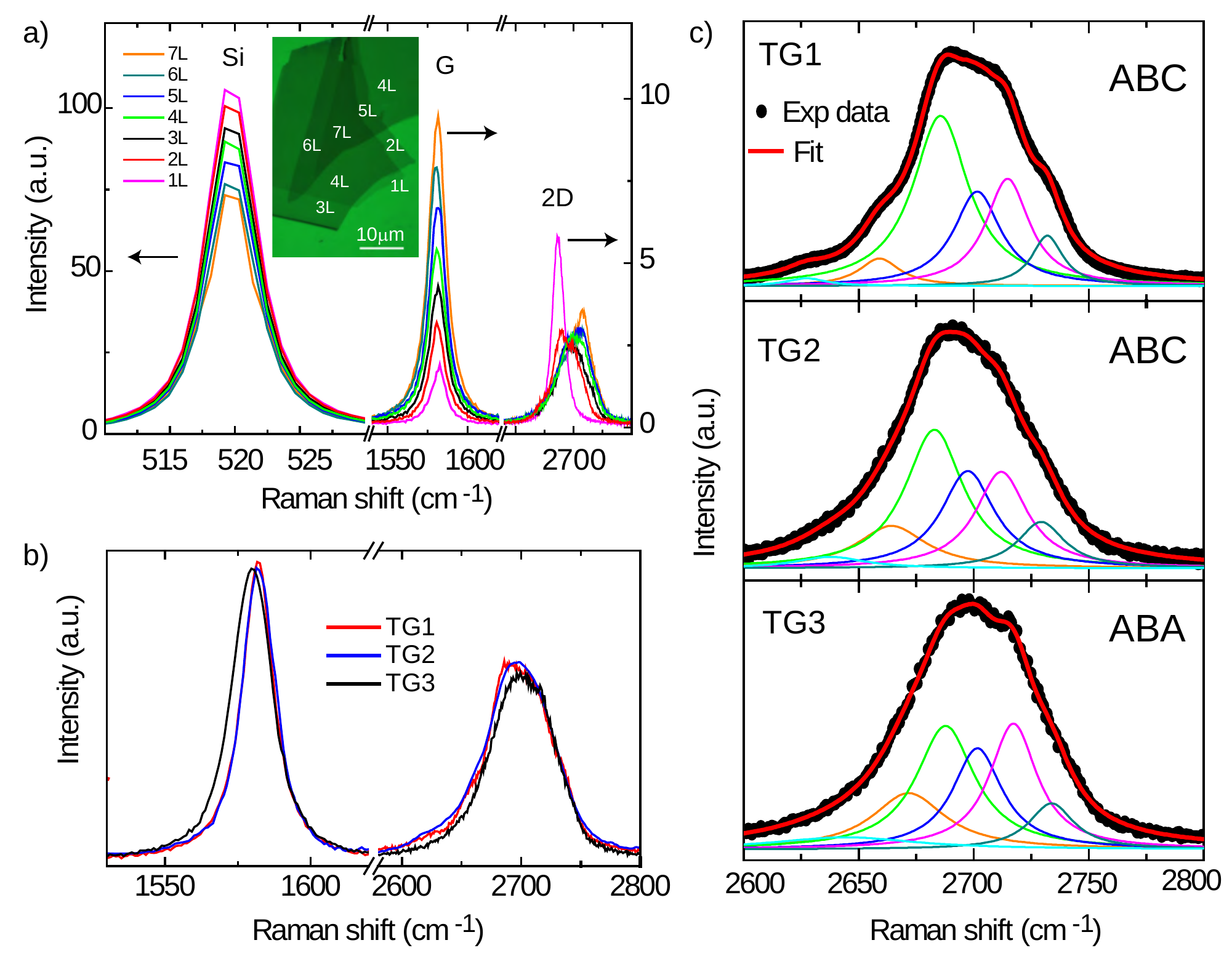}
\noindent{\caption{(a) Raman spectra for graphene samples with different number of layers. The inset shows the optical microscope picture of the flake containing up to 7 layers used for these measurements. The labels 1L to 7L indicate the number of layers. (b) Raman spectra of trilayer samples discussed in this work. (c) The 2D Raman band of graphene trilayers with ABC stacking (TG1 and TG2) and ABA stacking (TG3). The red lines are fits by 6 Lorentzian functions and the lines of other colors are the Lorentzian components of the fits.} \label{raman}}
\end{figure}

All our investigated graphene samples have been reliably identified as trilayers and their stacking order has been determined by means of Raman spectroscopy.
We used an excitation laser with a wavelength of 532 nm and a spot size of 1.5 $\mu$m in diameter. The Raman spectra of mechanically exfoliated graphene shows two peaks: the G band and the 2D (G${'}$) band at respectively 1580 cm$^{-1}$ and 2700 cm$^{-1}$ (Fig.~\ref{raman}(a)). The G band is due to the first-order Raman scattering by the double degenerate $E_{2g}$ phonon mode at the Brillouin zone center, while the 2D band originates from a second-order process, involving two intervalley optical phonons near the boundary of the Brillouin zone.\cite{Ferrari2006} The peak at 520 cm$^{-1}$ (labeled as Si) is due to the first-order Raman scattering by optical phonons of the Si substrate.

A reliable approach to count the number of layers ($N$) of FLG deposited on Si/SiO$_{2}$ substrates is based on the ratios of the intensities of the G peak and the Si peak, I$_{\text{G}}$/I$_{\text{Si}}$.\cite{Koh2011} As shown in Fig.~\ref{raman}(a), for a flake containing up to 7 layers, the intensities of the G and Si peaks clearly change with $N$. We find that I$_{\text{G}}$/I$_{\text{Si}}$ increases monotonically and discretely with $N$ due to an increase of the intensity of the G peak and a decrease of the intensity of the Si peak. Our findings are in agreement with recent observations, which attribute this behavior to enhanced absorption and Raman scattering of light by thicker graphene layers.\cite{Koh2011} In Figure \ref{raman}(b), we show the Raman spectra of all trilayers investigated here. These samples have G and 2D peaks of similar intensities and their I$_{\text{G}}$/I$_{\text{Si}}$ is consistent with the typical values found for trilayer graphene.

An accurate determination of $N$ for up to three layers is also possible from the 2D peak since its shape and position evolves with $N$ (see Fig.~\ref{raman}(a)). The 2D band is affected by the band structure of the material since it arises from a double-resonance process involving transitions among various electronic states. As trilayer graphene has three valence and three conduction bands, up to 15 electronic transitions can contribute to the 2D band.\cite{Malard2009} However, many of these different processes have very close energy separations and experimentally it is found that the minimum number of Lorentzian functions necessary to correctly fit the 2D mode of trilayer graphene is six.\cite{Malard2009,Lui} Consistently, Fig.~2(c) shows that for all the investigated trilayer graphene samples a good fit can be achieved with 6 Lorentzian functions.

Having determined the number of layers, we now address the stacking order in our trilayer samples. It has been recently demonstrated that an accurate and efficient method to characterize stacking order in FLG is based on the distinctive features of the Raman 2D peak.\cite{Lui} We find that TG1 and TG2 show a more asymmetric 2D peak than TG3, consistent with the reported differences between ABC and ABA stacking (see Fig.~\ref{raman}(c)).\cite{Lui} These differences in the 2D band feature are best captured by the Lorentzian components of their fits. In particular, the Lorentzians with the highest intensities - i.e., centered around 2685 cm$^{-1}$ (green) and 2715 cm$^{-1}$ (purple) - have very different intensities in the ABC samples (TG1 and TG2), whereas they have almost equal intensities in the ABA trilayers (TG3), in agreement with the observations reported in Ref.~[\onlinecite{Lui}].

\begin{figure}[t]
\includegraphics[width=7.5cm]{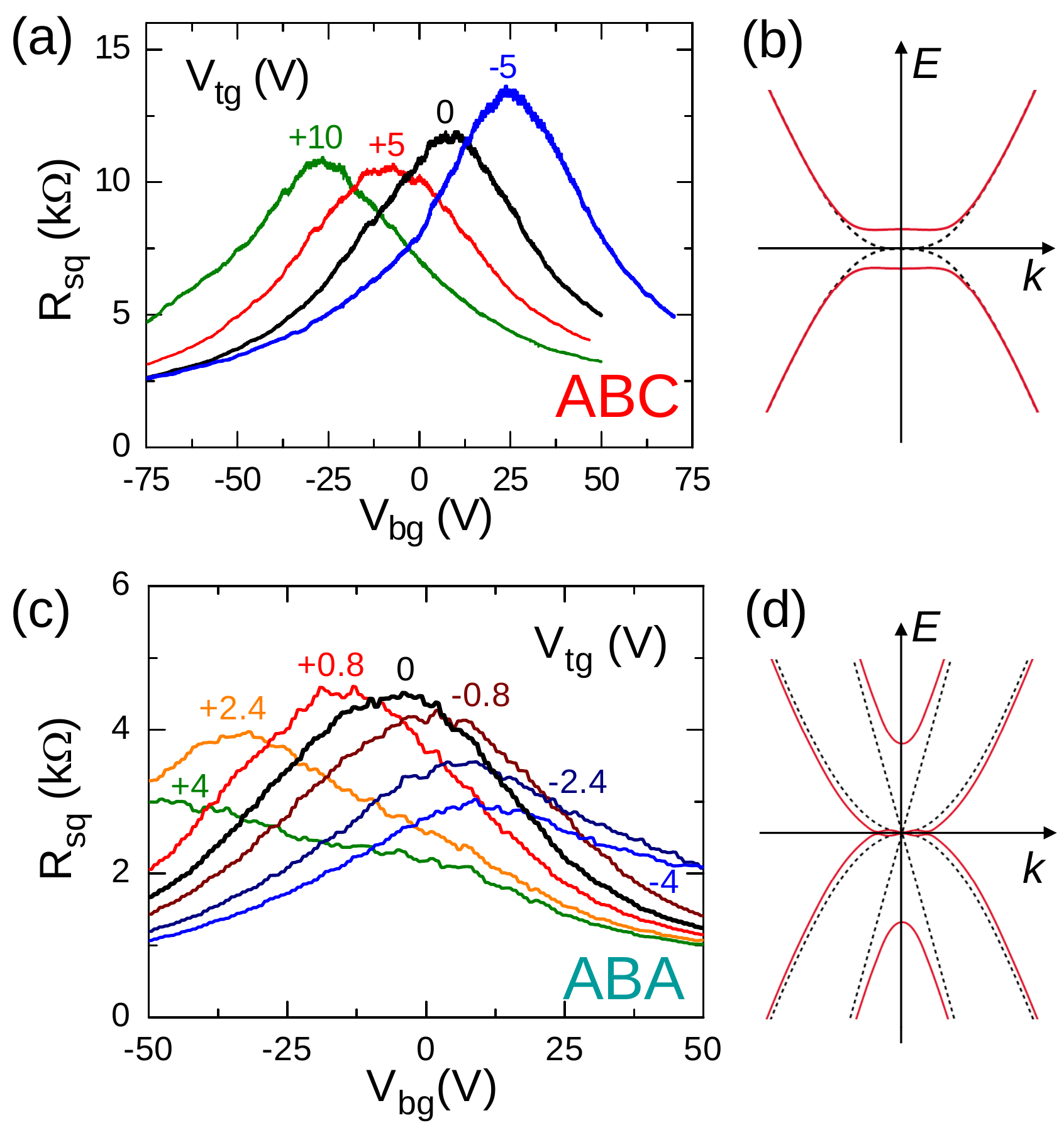}
\caption{(Color online) Square resistance $R_{\text{sq}}$ as a function of back-gate voltage for different fixed values of top-gate voltage at 4.2~K,
shown for (a) ABC trilayer and (c) ABA trilayer with thicknesses of top-gate dielectric of 90 and 15~nm, respectively.
Figures on the right present the schematic band structure of (b) ABC and (d) ABA trilayer graphene (considering only $\gamma_0$ and $\gamma_1$).
Application of a perpendicular electric field opens an energy gap for the ABC trilayer, while it results in a band overlap for the ABA trilayer.
Solid and dashed lines are with and without the external electric field, respectively.
}\label{fig:double}
\end{figure}

We now turn our attention to the transport properties of trilayer graphene in double-gated transistor structures.
This device geometry allows the independent control of the Fermi energy and the external perpendicular electric field $E_{\text{ex}}$ applied to the trilayers.
In particular, the $E_{\text{ex}}$ is given by $E_{\text{ex}}=V_{\text{tg}} / d_{\text{tg}} - V_{\text{bg}}/d_{\text{bg}}$ with $V_{\text{tg}}$ and $V_{\text{bg}}$ the top- and back-gate voltages respectively, and $d_{\text{tg}}$ and $d_{\text{bg}}$ the thicknesses of the top- and back-gate dielectric. Fig.~\ref{fig:double}(a) and (c) show the 2-terminal square resistance ($R_{\text{sq}}$) of trilayers with different stacking orders, measured for fixed values of $V_{\text{tg}}$ as a function of $V_{\text{bg}}$. In all cases $R_{\text{sq}}$ displays a maximum ($R_{\text{sq}}^{\text{max}}$) corresponding to the charge neutrality in the system. Clearly, the evolution of $R_{\text{sq}}^{\text{max}}$ with $E_{\text{ex}}$ is markedly different for the two stacking orders. For ABC trilayer $R_{\text{sq}}^{\text{max}}$ increases with increasing $E_{\text{ex}}$, whereas the opposite behavior is observed for ABA trilayer, i.e. $R_{\text{sq}}^{\text{max}}$ decreases with increasing $E_{\text{ex}}$. In both cases the position in $V_{\text{bg}}$ of $R_{\text{sq}}^{\text{max}}$ shifts linearly with $V_{\text{tg}}$, reflecting the changes in charge density induced by the two gates.

These results can be understood by the effect of the perpendicular electric fields on the band structure of ABA and ABC graphene trilayers.
Theory predicts that the interlayer asymmetry induced by the electric field opens an energy gap in the band structure of ABC trilayers (Fig.~\ref{fig:double}(b)),\cite{Guinea,Koshino1,Avetisyan}
whereas it causes a band overlap in ABA trilayers (Fig.~\ref{fig:double}(d)).\cite{Koshino3}
The electric-field tunable energy dispersion is a unique characteristic of few-layer graphene materials, and it paves the way to devices with unprecedented functionalities.
Recent experiments in double-gated bilayer transistors have demonstrated an on/off current ratio of 100 at room temperature.\cite{Avouris} On the other hand, very little is known experimentally on the electric-field tunability of the band structure of thicker few-layers and their stacking dependence.

\begin{figure}[b]
\includegraphics[width=8.6cm]{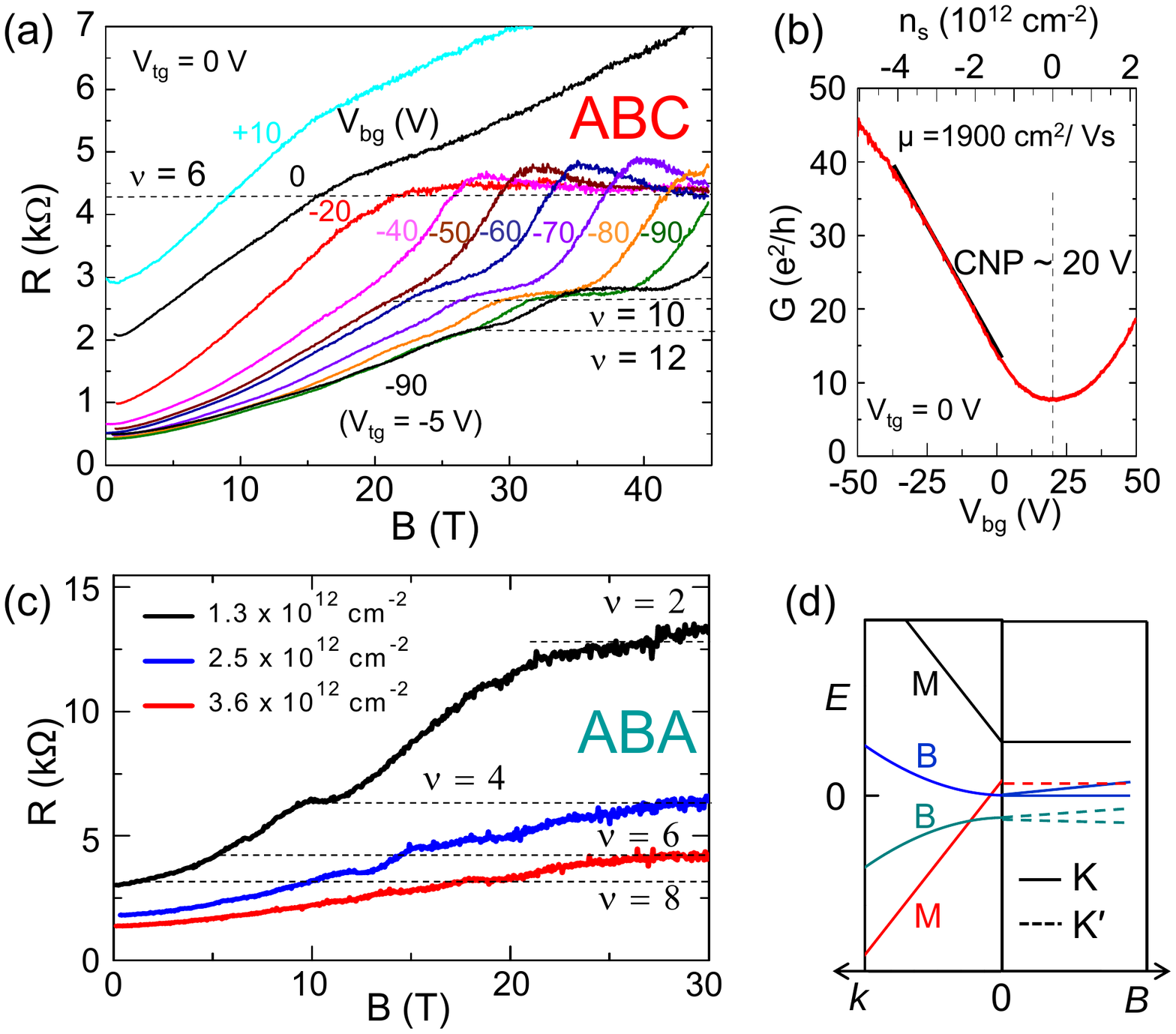}
\caption{(Color online) (a) Two-terminal magnetoresistance of the ABC-stacked trilayer at 4.2~K, shown for various back-gate voltages at $V_{\text{tg}}=$~0, except for a trace indicated.
QHE plateaus develop at $\nu =$~6,~10, and~12 (dashed lines). The small deviations from the dashed lines reflect the contact resistance of our device (TG1).
(b) Conductance $G$ of the ABC trilayer device (TG1) as a function of $V_{\text{bg}}$ (or carrier density $n_{\text{s}}$) at $V_{\text{tg}}= 0$. The mobility, $\mu$, is estimated from the linear $V_{\text{bg}}$ dependence of $G$ at large back-gate voltages.
(c) Magnetoresistance of ABA-stacked trilayer (TG3) shown for three different carrier densities at 4.2~K.
(d) Schematic low-energy band structure (left) and zero-energy LLs (right) of ABA trilayer graphene with all hopping parameters $\gamma_0-\gamma_5$ included.
The next-nearest layer couplings $\gamma_2$ and $\gamma_5$ shift the energy of monolayer-like (M) and bilayer-like (B) bands relative to each other, and also split zero-energy LLs into valleys.
}\label{fig:QHE}
\end{figure}

Fig.~\ref{fig:QHE} illustrates the effect of a perpendicular magnetic field on the transport properties of ABA and ABC trilayer graphene at $T =$~4.2~K. For ABC-stacked trilayer (TG1, $\mu \approx$~1900~cm$^{2}$V$^{-1}$s$^{-1}$) the 2-terminal magnetoresistance shows well-defined QHE plateaus at $\nu =$~6 and~10 for $B > 20$~T (Fig.~\ref{fig:QHE}(a)).
The filling factor $\nu = n_{\text{s}}\phi_0 B^{-1}$, where $\phi_0$ is the flux quantum, matches well with the carrier density
$n_{\text{s}}=\alpha(V_{\text{bg}}-V_{\text{CNP}})$ calculated using $\alpha=7.2 \times 10^{10}\text{cm}^{-2}\text{V}^{-1}$.
The observed plateaus are expected from the 3-fold degenerate zero-energy LLs of the ABC trilayer graphene ($E_n \propto B^{3/2} \sqrt{n(n-1)(n-2)}$)
with 4-fold spin and valley degeneracy. We find QHE plateaus only away from the charge neutrality point (CNP) located at $V_{\text{CNP}}\sim$~20~V for $V_{\text{tg}}= 0$ (Fig.~\ref{fig:QHE}(b)).
An additional plateau develops at $\nu =$~12, rather than at the expected $\nu =$~14, upon further increase of $E_{\text{ex}}$ (for example at $V_{\text{bg}}=$~-90~V with $V_{\text{tg}}=$~-5~V). This observation suggests lifting of the valley degeneracy induced by the interlayer potential asymmetry,\cite{Koshino2} imposed by the top and back gates.

The ABA-stacked trilayer device (TG3, $\mu \approx$~1100~cm$^{2}$V$^{-1}$s$^{-1}$) develops QHE plateaus at $\nu =$~2,~4,~6, and~8 with a step of $\Delta\nu =$~2 (Fig.~\ref{fig:QHE}(c)).
This observation is consistent with a recent theoretical prediction which includes the complete set of hopping parameters up to the next-nearest layer couplings $\gamma_2$ and $\gamma_5$.\cite{Koshino4}
This extended model predicts relative energy shifts of the monolayer-like and bilayer-like LLs in the ABA trilayer and a valley split of the zero-energy LLs by the $\gamma_2$ and $\gamma_5$.
As a result, the 12-fold zero-energy levels (4 and 8 zero-energy levels from the monolayer-like and the bilayer-like subbands, respectively) split into 6 different energies with twofold spin degeneracy (Fig.~\ref{fig:QHE}(d)), leading to the QHE plateaus at filling factor intervals of $\Delta\nu =$~2.
In addition, the presence of the external electric field generally splits the valley degeneracy of the LLs by the induced interlayer asymmetry.\cite{Koshino2}
As opposed to the case of ABA trilayer, the electric-field-induced valley splitting is expected to be smaller for the inversion-symmetric ABC trilayer.
Therefore, the 4-fold spin and valley degeneracy is retained for the ABC trilayer device, resulting in QHE plateaus at $\nu =$~6 and~10.
Under the large external electric field, however, the valley splitting leads to the QHE plateau at $\nu =$~12.

During the writing of our manuscript, we became aware of preprints dealing with QHE in ABA\cite{Taychatanapat} and ABC trilayers.\cite{Zaliznyak,Bao,Kumar}
In Ref.~[\onlinecite{Taychatanapat}], QHE plateaus in the ABA trilayer are observed at $\nu = \pm2, \pm4$,~-6 but not at $\nu =~+6$.
The absence of a plateau at $\nu = +6$ is attributed to LL crossing.
As pointed out in the Ref.~[\onlinecite{Taychatanapat}], actual plateaus developing in the ABA trilayer can depend on a specific $B$ (or in our case $V_{\text{bg}}$)
where measurements are performed, due to the LL crossing between the monolayer-like and the bilayer-like subbands.
For the ABC trilayer, Refs.~[\onlinecite{Zaliznyak}] and [\onlinecite{Kumar}] report QHE plateaus at $\nu = \pm6, \pm10, \pm14...$, consistent with our results except the plateau at $\nu = 12$,
whereas Ref.~[\onlinecite{Bao}] observed rather unexpected plateaus at $\nu = \pm9, \pm18$, and -30.

In summary, we have investigated transport properties of trilayer graphene with different stacking order.
Samples with ABA and ABC stacking differ characteristically in the sequence of quantum Hall plateaus, in agreement with recent theory.
The stacking order provides an additional degree of freedom to tune the electronic properties of trilayer graphene,
combined with the interlayer asymmetry controlled by top and back gates.

We thank A.~Morpurgo for suggesting this experiment and D.~Weiss for continuing support.
We acknowledge financial support from the Deutsche Forschungsgemeinschaft within GRK~1570,
EuroMagNET (EU Contract no.~228043), EPSRC (Grant no.~EP/G036101/1 and no.~EP/J000396/1), the Royal Society Research Grant 2010/R2 (Grant no.~SH-05052),
Grant-in-Aid for Young Scientists A (no.~20684011), ERATO-JST (080300000477), Special Coordination Funds for Promoting Science and Technology (NanoQuine), JST Strategic International Cooperative Program and MEXT Grant-in-Aid for Scientific Research on Innovative Areas (21102003).

\end{document}